# High-sensitivity electrochemical dual-QCM for reliable three-electrode measurements

*Dávid Tóth[1,2], Manuel Kasper[1], Ivan Alic[1,2], Mohamed Awadein[1,2], Andreas Ebner[2], Doug Baney[3], Georg Gramse[1,4] and Ferry Kienberger[1]*

1 Keysight Technologies GmbH, Linz 4020, Austria; manuel.kasper@keysight.com; mohamed.awadein@keysight.com

2 Applied Experimental Biophysics, Johannes Kepler University, Linz 4020, Austria; david.toth@jku.at; ivan.alic@jku.at; andreas.ebner@jku.at

3 Keysight Technologies, Santa Clara 95052, USA; doug_baney@keysight.com

4 Molecular Biophysics and Membrane physics, Johannes Kepler University, Linz 4020, Austria; georg.gramse@jku.at

*Correspondence: ferry_kienberger@keysight.com

**Keywords**

Quartz crystal microbalance, electrochemistry, underpotential deposition, dual QCM, ECQCM

## Abstract

Electrochemical quartz crystal microbalance (EC-QCM) is a versatile gravimetric technique that allows for parallel characterization of mass deposition and electrochemical properties. Despite its broad applicability, simultaneous characterization of two electrodes remains challenging due to practical difficulties posed by the dampening from fixture parasitics and the dissipative medium. In this study, we present a dual electrochemical QCM (dual EC-QCM) that is employed in a three-electrode configuration to enable consequent monitoring of mass deposition and viscous loading on two crystals, the working electrode (WE) and the counter electrode (CE). A novel cor-rection approach along with a three standard complex impedance calibration is employed to overcome the effect of dampening while keeping high spectral sensitivity. Separation of viscous loading and rigid mass deposition is achieved by robust characterization of the complex impedance at the resonance frequency. Validation of the presented system is done by cyclic voltammetry characterization of Ag underpotential deposition on gold. The results indicate mass deposition of 412.2 ng for the WE and 345.6 ng for the CE, reflecting a difference of the initially present Ag adhered to the surface. We also performed higher harmonic measurements that further corroborate the sensitivity and reproducibility of the dual EC-QCM. The demonstrated approach is especially intriguing for electrochemical energy storage applications where mass detection with multiple electrodes is desired.

# 1. Introduction

Quartz crystal microbalance (QCM) is a well-established technique in the field of gravimetry with a number of applications in aerosol detection, thin film deposition, biology and electrochemistry [1–5]. In various fields of electrochemistry such as such as lithium-ion battery characterization, electrodeposition or corrosion, QCM studies have been reported in the recent years focusing on characterization of a single resona-tor [6,7]. However, a universal understanding of intrinsic electrochemical processes, particularly in the field of battery research, often requires characterization in real or close to real conditions with simultaneous monitoring of multiple electrodes. In this study, we present an electrochemical QCM setup employing two quartz crystals, re-ferred to as dual EC-QCM, which allows for concurrent characterization of mass dep-osition and dissipative loading of two quartz sensors in a three-electrode configuration.

QCM relies on the changes of the resonant piezo crystal oscillation upon deposition of mass or property changes of the surrounding medium. A common way to char-acterize the oscillation is by measurement of the QCM electrical response. Sauerbrey was the first one to describe the relation between the shift of the characteristic resonance frequency and the deposited mass [8]. While the Sauerbrey-relation considers the crystal mechanical properties in order to characterize the deposited mass of a thin, rigid layer, it does not account for changes due to an introduced dissipative loading. Further considerations for operation in dissipative conditions were described by Kanazawa, who has shown that the effect of liquid loading on the characteristic resonance frequency is:

$$\Delta f = f_0^{3/2} \left(\frac{\rho \eta}{\pi \mu_q \rho_q}\right)^{1/2},$$

where Δf is the resonance frequency change, f0 is the resonance frequency of the un-perturbed crystal, ρ and η represents the liquid density and viscosity respectively, µq is the quartz shear modulus (~2.947 x 10¹¹ g cm-1 s2) and ρq is the quartz density (~2.651 g cm-3) [9]. This relation was further developed by Martin et al, who has shown that separation of liquid loading and mass deposition is possible with the extensive characterization of the crystal complex impedance response [10]. In their study they showed that in the case of a thin rigid deposited layer in liquid conditions the following relations can be applied:

$$\Delta f = - \frac{2f_0^2}{N\sqrt{\rho_q \mu_q}} \left[ \rho_s + \left( \frac{\rho \eta}{4\pi f_0} \right)^{1/2} \right],$$

where N indicates the order of the resonance harmonic and ρs is the deposited mass density. As the reactance reaches zero at the resonance frequency, the absolute value of the complex impedance is dominated by the real part value:

$$Z_{Re,min} = \frac{\eta_q}{\rho_q C_1} + \frac{1}{N\pi C_1} \left( \frac{\rho \eta}{\pi f_0 \rho_q \mu_q} \right)^{1/2},$$

with ZRe,min the minimum of the real part of the complex impedance at the resonance frequency, ηq the effective quartz viscosity and C1 the motional capacitance of the crystal. Therefore, a deposited rigid layer also contributes to the stored kinetic energy in the oscillation. The motional capacitance can be understood from the Butter-worth-van-Dyke (BVD) model that describes the crystal response as an RLC series resonator (R1, L1, C1) with an additional parallel geometrical capacitance (C0). Martin et al. also proposed additions to the BVD circuit to account for the liquid and mass in-teractions with an inductor (LL) and a resistor (RL) representing the liquid loading, and an additional inductor (Ls) as the loading of the deposited solid film. The described study pointed out the importance of characterization of the unperturbed crystal prop-erties in order to separate the liquid and mass loading effects. This can only be achieved after reliable calibration of the measurement system by removal of parasitic impedances arising from the fixturing such as the introduced capacitance from con-nectors or additional resistance and inductance from the wiring. The above-mentioned approaches allowed for complex investigations in liquid such as description of molec-ular interactions, state-of-charge characterization in lead-acid batteries and underpo-tential deposition study on gold electrodes [1,6,11–14]. Technical capabilities of QCM were further extended by measuring at higher harmonic frequencies of the crystal os-cillation. Such characterization is possible as the electrical perturbation of a QCM re-sults in the excitation of the odd number harmonics [12]. The higher harmonics allow for robust separation of rigid mass deposition and dissipative loading as $\Delta f$ varies with $N\rho s$, while $\Delta f$ and $Z_{Re,min}$ vary with $(N\rho\eta)^{1/2}$ [10].

In this study, the aforementioned approaches are employed to validate a dual EC-QCM configuration by characterization of underpotential deposition of Ag on two quartz sensors simultaneously. As the ability to resolve mass deposition in the range of nanograms requires high sensitivity and stability, a complex impedance correction procedure is employed to overcome practical difficulties, such as liquid dampening. The presented correction workflow allows separation of liquid loading properties and effects of rigid mass deposition with high sampling rate frequency tracking for both QCMs in dissipative medium. Finally, multi-harmonic measurements using both working and counter electrode crystals are performed to corroborate the findings.

# 2. Materials and methods

## 2.1. Dual EC-QCM connection scheme

The connection scheme for the dual EC-QCM is shown in Figure 1 and detailed in Fig-ure S1. A Keysight E4990 impedance analyzer (Keysight Technologies, Santa Rosa, CA, USA) with a Keysight 42941A impedance probe is used to perform the resonance fre-quency measurements of the two QCM crystals. A Keysight B2902A (Keysight Tech-nologies, Santa Rosa, CA, USA) source measure unit (SMU) is utilized for the voltage sweep and corresponding current measurement. Four-terminal connections are em-ployed to enable a parasitic free, three-electrode measurement consisting of the two QCM crystals acting as counter electrode (CE) and working electrode (WE), and a Ag wire acting as the reference electrode (RE). The fixture consists of a two bias-tee con-figuration to separate the transmission line of the RF and DC signals. With the addition of the bias-tees, the circuitry can be represented with the equivalent circuit model shown in Figure 2a and the schematic depiction shown in Figure 2b. The fixture, the motional branch and the geometrical capacitance of each of the two resonators (Crys-tal 1 and Crystal 2) as well as the mass and liquid loading are represented by the or-ange, grey, yellow, green and blue squares, respectively. Cb and Lb describe the capaci-tance and inductance of the bias-tees. C1, R1, and L1 relate to the motional branch of Crystal 1 and describe the capacitance, resistance, and inductance, respectively. Parallel to the motional branch is C01 describing the geometrical capacitance, and parallel to Crystal 1 is Crystal 2 with the same elements (subscript '2'). LS describes the mass loading, and RL and LL the resistive and inductive liquid loading, respectively. The ex-tracted equivalent circuit parameters of the unperturbed, parallelly connected Crystal 1 (9 MHz resonance frequency) and Crystal 2 (10 MHz resonance frequency) con-nected in parallel are given in Table 1. The observed differences in the equivalent circuit parameters reflect the resonance frequency and sensitivity difference of the two crystals.

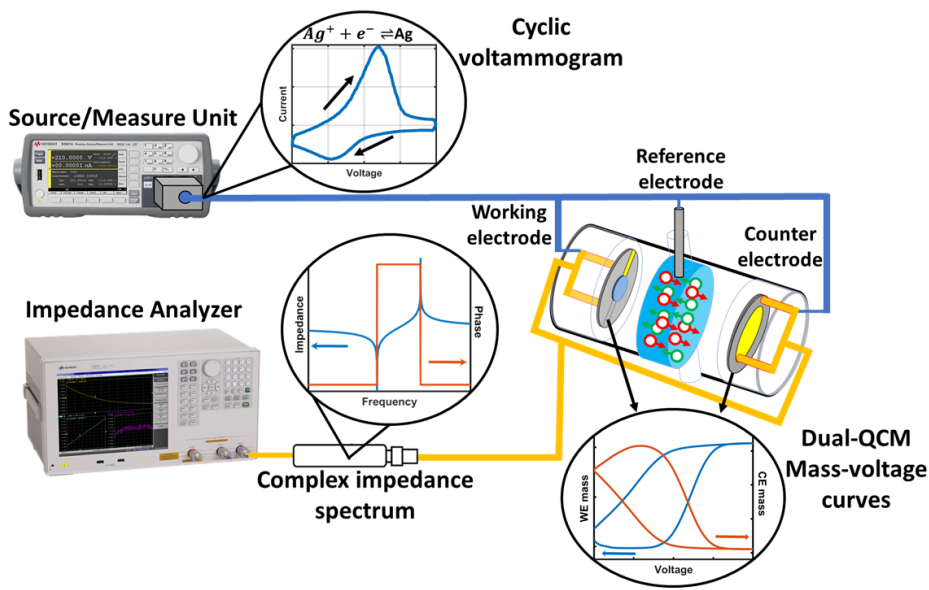

*Figure 1. Schematics of the dual EC-QCM setup, where the two quartz resonators act as the working electrode (WE) and counter electrode (CE) in a three-electrode configuration. First, the impedance analyzer is used to characterize the unperturbed crystal impedance prior to the measurement. Then, the source/measure unit performs the cyclic voltammetry measurement with a Kelvin-connection in a three-electrode configuration, including the reference electrode. Thereby, in parallel to the CV, the complex impedance spectra of both QCMs are characterized subsequently allowing the calculation of mass deposition on both electrodes.*

## 2.2. Dual EC-QCM cell

A homebuilt QCM cell (Johannes Kepler University, Linz, Austria) was modified by the addition of separate crystal holders on each side of the liquid chamber in order to accommodate two resonator crystals. Figure S2 shows a picture of the dual EC-QCM cell. The electrical connections to the QCMs are realized via spring contacts that are directly connected to BNC adaptors at each side of the cell, to allow for parasitic free measurements at high frequency. The ~1 ml volume liquid compartment is directly connected to two opening at opposite sides, enabling continuous flow of electrolyte. The liquid cell compartment prepared from polypropylene assures chemical stability with the used electrolyte. Two quartz crystals from QUARTZ PRO AB (Jarfalla, Sweden) with resonance frequencies of 9 and 10 MHz and 14 mm Cr/Au electrodes on each side were used.

## 2.3. Electrochemical test system

For validation of the dual EC-QCM Ag underpotential deposition (UPD) on Au surface was chosen as the model electrochemical system. Ag UPD allows for precise deposition and removal of mass by controlling the applied voltage [15–17]. Another benefit of metal UPD is that for thin layer deposition, the Sauerbrey gravimetric conditions hold allowing for separation of liquid and mass loading effects [8]. The measurements were performed in aqueous 0.2 M $H_2SO_4$, $10^{-3}$ M $Ag^+$ solution. $Ag_2SO_4$ (Sigma Aldrich), $H_2SO_4$ and distilled water were used to prepare the electrolytes. In the electrochemical cycling a Ag wire (Sigma Aldrich) was employed as the reference electrode. Before measurements, the crystals and the reference electrode were washed with distilled water, isopropyl alcohol and dried with nitrogen flow to ensure removal of organic contaminants. In order to minimize the effect of the AC perturbation on the electrochemical reactions, a small 50 mV peak to peak amplitude was applied for the complex impedance characterization. We note that in the applied electrochemical system upon liquid insertion, a small change in mass is expected even in open circuit conditions due to possible trapped charges allowing for reduction of Ag+ onto the electrode. The deposition of this initial mass inevitably results in a small error when separating mass and loading effects during the electrochemical cycling. However, because the magnitude of this initial mass is very low, we assume that the frequency response of the crystal upon insertion in the liquid represents changes only due to liquid loading.

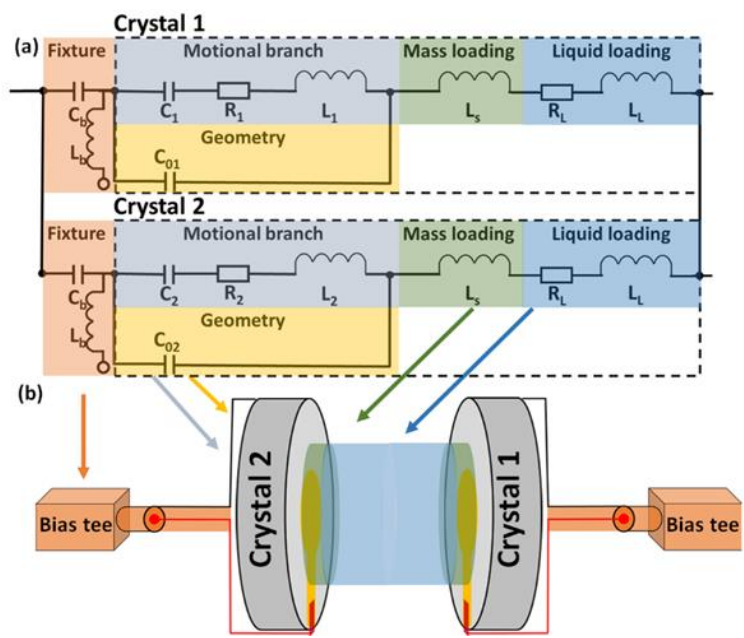

*Figure 2. Circuit model of the dual EC-QCM. (a) Equivalent circuit model of the two crystals in parallel including separate rigid mass and liquid loading elements as well as series capacitors (Cb) and parallel inductors (Lb) representing the bias-tees. (b) Schematic depiction of the two crystals with the two bias tees, as well as the physical phenomena represented by the equivalent circuit model elements.*

**2.4. Dual EC-QCM software control**

Custom written MATLAB scripts using SCPI commands were employed to control the instrumentation. The control procedure includes the following steps. Firstly, the resonance frequencies of both crystals at the selected harmonics are identified. Sec-ondly, a broadband extraction of the complex impedance centered around the reso-nance frequency is performed for accurate measurement of the resonator parameters. Thirdly, the cyclic voltammogram (CV) is set up on the SMU with the user-defined parameters. Finally, the DC voltage sweep is initialized by the SMU with subsequent frequency sweeps done by the impedance analyzer at the identified resonance fre-quencies. During the measurements both the resonance frequency as well as the real and imaginary part of the complex impedance are monitored for a complete charac-terization of the crystal response. Considering previous studies that have shown the advantage of tracking the zero reactance (zero-phase) point instead of the maximum of the conductance in dissipative medium, in our implementation we applied the ze-ro-phase point tracking [18]. An overview of the measurement workflow during the CV is shown in Figure S3.

## 2.5. Complex impedance correction

The effect of the combined dampening of the fixture and the dissipative medium is shown in Figure S4, where we plot the magnitude and phase of the complex imped-ance of a single crystal in four different scenarios. First, the response of the QCM crys-tal with direct connection to the impedance probe is shown, then the crystal response in air with the fixture, then the crystal response in liquid with the fixture in place and finally under the latter conditions but with a particular correction described below. Overall, the shown spectra clearly indicate that QCM measurements in liquid with a fixture make the frequency tracking challenging. To compensate for the dampening, we employed a complex impedance correction step (Figure 3) which is performed be-fore the CV and the frequency and resistance tracking procedure. The impedance correction uses an open-standard $Y_{open}=1/Z_{open}$ described by a par-allel capacitance ($C_{open}$) and resistance ($R_{open}$)

$$Y_{open} = 1/R_{open} + j\omega C_{open},$$

where j is the imaginary unit and ω is the angular frequency [19]. The removal of this impedance component on a broad frequency range can be regarded as a removal of stray impedance components such as C0 and the fixture contributions Rf and Cf that otherwise distort the complex impedance response of the resonator [12]. By subtraction of Zopen from the measured impedance, a more pronounced motional resonance response is achieved as shown in Figure S4. Additionally, as a post-processing step for accurate dissipative loading separation, the uncorrected impedance spectra was recalculated. In Figure S5, we show the comparison of the distorted and the recalculated resonance frequency shift and resistance change over multiple cycles of the CV. The resonance frequency shows a significant ~80 Hz difference at certain voltages, whereas the resistance change for the two cases shows variation of a few Ohms and an overall different behavior over time. The presented impedance correction in combination with the recalculation procedure enables reliable frequency tracking by measuring the complex impedance at a small frequency range, therefore fast sampling rates can be employed for electrochemical systems that can result in larger mass deposition or viscoelastic property change over a short period of time.

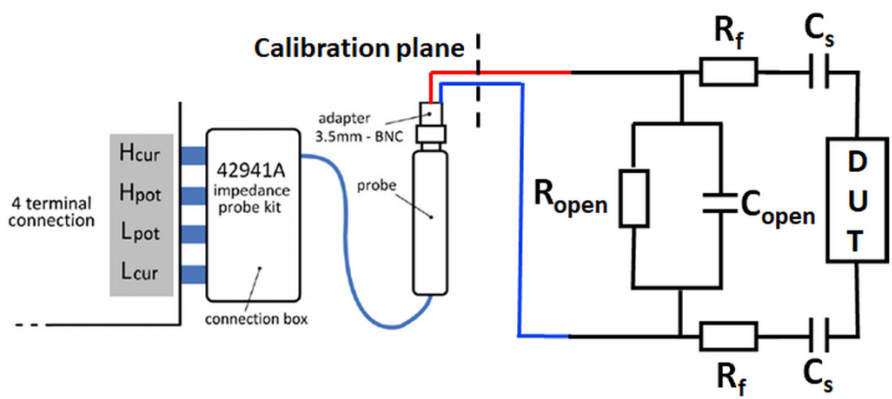

**Figure 3.** Complex impedance correction procedure allowing high sampling rate frequency tracking in dis-sipative media. Equivalent circuit model of the setup during the measurements including the open-standard (Copen, Ropen), and Cf and Rf representing the fixture stray impedance.

**2.6 Higher harmonics**

Further demonstration of the versatility of the presented dual EC-QCM setup is shown in Figure S6, where we plot results of the third harmonic measurements for both crystals. Application of higher harmonics is advantageous for measurements re-quiring higher sensitivity [12]. The frequency change due to mass deposition shows a linear dependence against the order of the oscillation harmonic as shown in Equation 3, thus at higher harmonics a larger overall frequency shift is observed. Although more noise is present at higher frequencies due to the larger influence of environmental parameters, the resolved mass calculation shows good agreement with the first harmonic data. Overall, the application of the impedance correction and loading separation shows good repeatability and confirm the validity of the workflow also for higher frequency measurements.

# 3. Results and Discussion

## 3.1. Dual EC-QCM sensitivity and stability

The ability to resolve mass deposition in the range of nanograms requires high sensitivity and stability, therefore performance characterization of the presented dual EC-QCM configuration is of key importance. Figure 4a shows that the resonance fre-quency of both crystals changes only by 0.5-1 Hz over the course of 15 minutes when measured in ambient conditions. This frequency stability enables accurate characteri-zation of mass deposition of the electrochemical model system as a resonance fre-quency shift of tens of Hz is expected for the employed metal concentration and volt-age sweep rate. In order to assess the resolution of the dual EC-QCM, in Figure 4b we show the mass against frequency shift curves for both crystals that were recorded during the UPD of Ag in aqueous 0.2 M $H_2SO_4$. The calculation of deposited mass was done after the impedance correction pro-cedure and the mass and liquid loading separation (see Methods section). The slopes of the two curves represent the sensitivity of the two QCMs, resulting in 1.4 ng/Hz and 2.7 ng/Hz for crystal 1 and 2, respectively. The observed difference between the sensi-tivity of the two crystals reflects the difference in the fundamental resonance frequen-cies as the 10 MHz crystal is more sensitive to mass changes compared to the 9 MHz crystal. The smallest frequency shift identified between two adjacent frequency points is 19 mHz for crystal 1 and 38 mHz for crystal 2 at a sampling rate of ~0.5 Hz. These frequency differences correspond to a mass change of 27 pg and 103 pg, or $1.5 \times 10^{11}$ and $5.7 \times 10^{11}$ number of Ag atoms on the complete electrode surface. These values represent the empirical rigid mass resolution of the dual EC-QCM indicating an exceptional sensitivity allowing high precision measurement of mass deposition in liquid medium for the two QCMs.

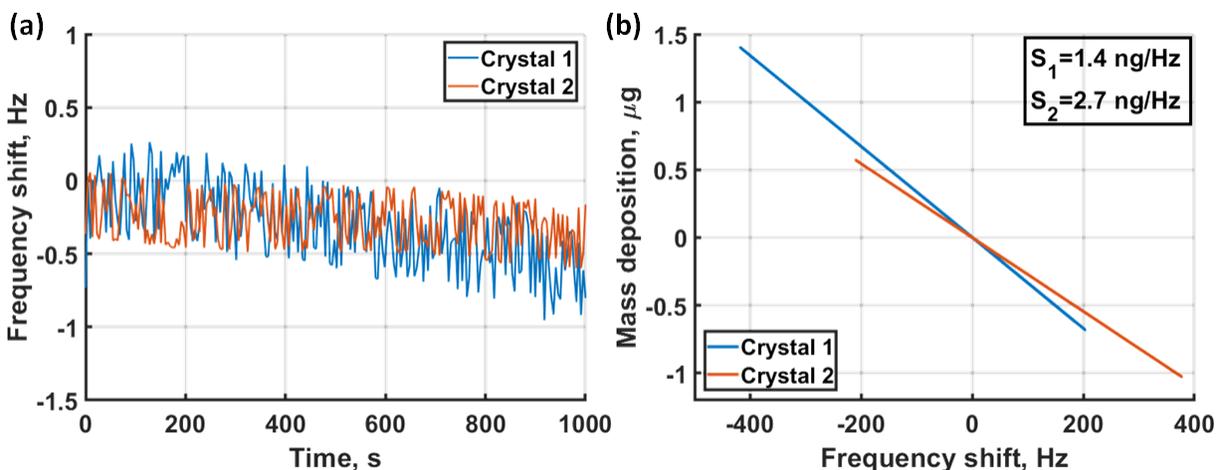

*Figure 4.* Stability and sensitivity analysis. (a) Resonance frequency stability of both crystals in ambient conditions. (b) Sensitivity analysis showing Δm against Δf curves of both crystals, indicating sensi-tivities of 1.4 and 2.7 ng/Hz respectively. At a sampling rate of 0.5 Hz the minimum frequency shifts of 19 mHz and 38 mHz are measured, resulting in a rigid mass resolution of 27 pg and 103 pg.

### 3.2. Electrochemical characterization

In order to characterize the electrochemical test system, in the first step we per-formed combined CV and QCM measurements in aqueous 0.2 M H2SO4 solution with-out the presence of Ag ions. As shown in Figure S7 no change in the crystal resonance frequency was observed during a CV, indicating that effects such as electric double layer formation or surface adhesion of ionic species are negligible. This is in agreement with previous studies that have shown only minor motional resonance frequency shifts of a few Hz which are due to the formation and decrease of electrical double layer in the vicinity of the electrode surface [20]. In the second step we identified the redox peaks for Ag UPD by using two different concentrations of Ag ions in the CV as shown in Figure S8. A clear increase of the peak currents at low potentials was ob-served after increasing the concentration of Ag in the electrolyte solution, indicating that the observed peaks correspond to the Ag UPD reaction peaks.

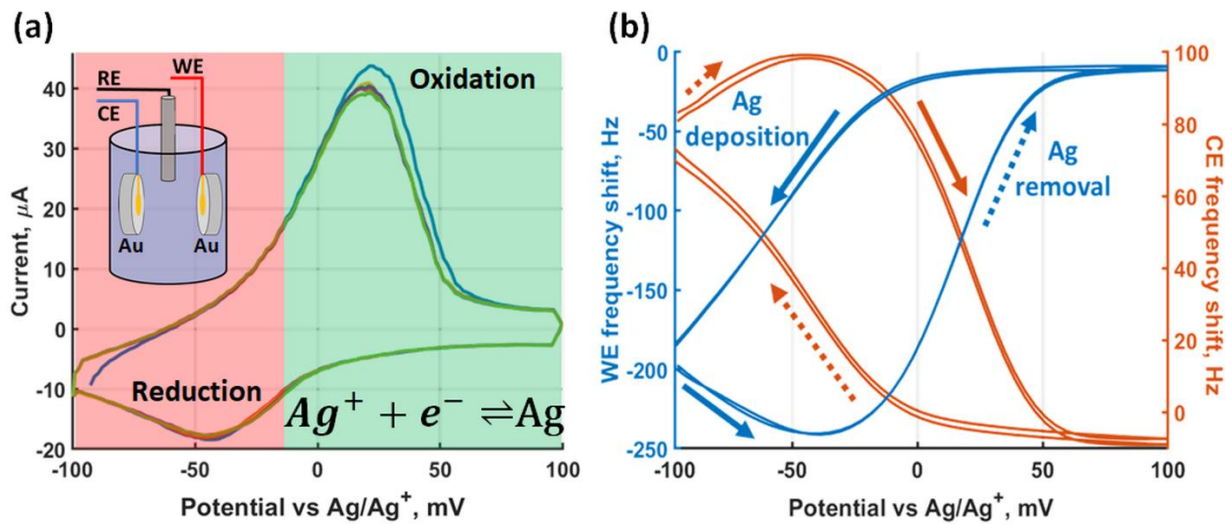

*Figure 5.* Dual EC-QCM raw data. (a) Cyclic voltammogram of Ag underpotential deposition in aqueous 0.2 M H2SO4, 10-3 M Ag2SO4 solution in a three-electrode configuration. (b) Relative frequency shift of the WE QCM and the CE QCM during the voltammogram. The solid arrows represent the di-rection of the frequency shift during Ag deposition while the dotted arrows correspond to the frequency shift during Ag removal.

In the third step, after the identification of the redox peaks, combined dual EC-QCM measurements were performed with 10-3 M Ag2SO4 concentration on the same voltage range at a scanning rate of 0.3 V/s. Figure 5a shows the recorded CV. The Ag oxidation peak is observed at ~15 mV and the reduction peak at -45 mV, resulting in a ~60 mV peak separation indicating electrochemical reversibility.

Simultaneous recording of the crystal resonance frequencies results in the response shown in Figure 5b, where we plot the relative frequency shift of the WE and the CE against the applied potential. The frequency shifts are relative frequency changes recorded after keeping the system at 0 V potential for one second. During the CV, the potential was swept from -100 mV to 100 mV. Starting at -100 mV, the WE frequency first exhibits a small drop at the initial negative voltages, indicating an uptake of mass (reduction of $Ag^+$-ions), then a sharp increase upon reaching positive potentials due to oxidation of Ag at the electrode surface, indicating a reduction of mass. In the reverse direction going from 100 mV to -100mV, sweeping towards negative potentials results in a reduction of $Ag^+$ onto the WE, therefore an uptake of mass. In contrast, for the CE, first, we observe a removal of initially present Ag at negative potentials, then a deposition of $Ag^+$ onto the gold surface at positive potentials and finally, a removal of Ag towards negative potentials in the return direction. The observed smaller magnitude of the CE resonance frequency shift compared to the WE can be explained by the lower sensitivity of the 9 MHz crystal and the CE-RE potential difference. Direct application of the Sauerbrey relation to this data results in a mass change of 409.1 ng for the WE and 336.7 ng for the CE (Figure S9). The difference in magnitude of the deposited mass could be partially attributed to the imbalance of the initially present Ag on the electrode surfaces and the contribution of a changing liquid loading on the resonance frequency. The presence of a dissipative liquid load would indicate that the Sauerbrey relation cannot be applied on its own to fully characterize the resonance frequency shift. The effect of the liquid loading on the mass changes is further explored in the following section.

**3.3. Loading separation for accurate mass quantification**

In order to assess the liquid loading contribution, the complete complex impedance response of the crystal resonance was investigated. Martin et al. have described that by characterization of the real part of the impedance response, separation of the liquid and the mass loading is possible [1]. Figure 6a shows the absolute value of the complex impedance versus the frequency at two arbitrary points during CV. The absolute value of the complex impedance is dominated at the motional resonance frequency by the mo-tional resistance.

Therefore, it is clear from the comparison that at different applied potentials a motional resonance frequency shift and a resistive load change are resolved. The resistance change over time for both crystals is in the range of 1-2 Ohms during the CV as shown in Figure 6b, indicating also a periodic change of the dissipation over the cycles. Equations 2 and 3 allow us to calculate the liquid loading (ρη) after a robust characterization of the unperturbed crystal response. This requires high confidence in the extracted unperturbed C1 motional capacitance and resonance frequency f0. This was achieved by a three-standard open-short-load calibration at the plane of the impedance probe and a subsequent complex impedance correction as described in the Methods section.

In Figure 6c we show the calculated frequency shift due to changes in the ρη liquid term as well as the separated rigid mass loading. It shows that most of the resonance frequency change is due to the rigid mass, which is responsible for a 368.8 Hz shift in the first cycle in the case of the CE, while only a 20-30 Hz variation is caused by the liquid loading. By applying both the impedance correction procedure and the frequency shift separation, the calculated mass deposition is shown in Figure 6d with respect to the applied potential. On the WE 412.2 ng Ag is deposited and on the CE 345.6 ng is deposited in the first cycle. While the magnitude of the deposited mass in a cycle shows only a difference of ~10 ng compared to the values extracted from the Sauerbrey equation, this difference reaches 25 ng at certain parts of the CV, introducing a considerable error. The results further indicate that the difference in the deposited mass on the WE and CE is likely due to the difference in the initially present Ag on the electrode surface. Overall, the separated frequency shift contributions prove the validity of the presented correction workflow for the dual EC-QCM configuration.

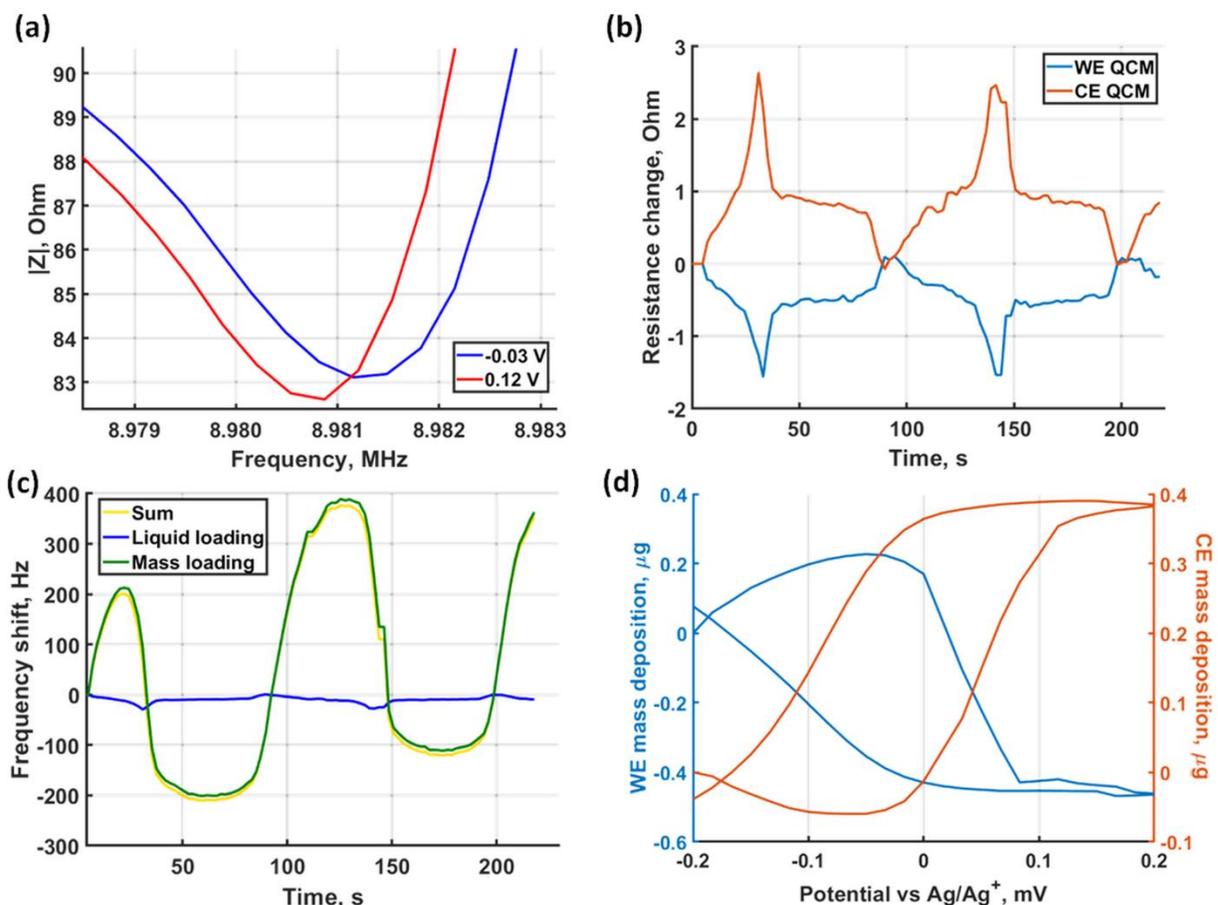

*Figure 6. Dissipative changes in the crystal oscillation. (a) QCM response at two arbitrary timepoints during CV. (b) Change of resonator resistance during CV. (c) Resonance frequency shift of the CE separated into dissipative loading and rigid mass deposition. (d) Mass change of both crystals calculated with liquid loading correction against the applied potential.*

## 4. Conclusion

In this study we presented a dual EC-QCM setup with three-electrode configuration that allows for robust and sensitive characterization of mass deposition and dissipative loading on the WE and the CE in parallel to cyclic voltammetry. A crystal stability of <1 Hz was demonstrated in ambient conditions. Along with a three standard complex impedance calibration an additional impedance correction procedure was implemented. This allows for high sensitivity measurements overcoming the dampening effects of both the system parasitics and the dissipative medium. Therefore, the dual EC-QCM approach facilitates reliable frequency tracking of highly dampened resonance characteristics. Underpotential deposition of Ag onto gold electrodes was employed to validate the versatility of the proposed setup. By applying data analysis introduced in previous studies we separate effects of rigid mass deposition and viscous loading. The quantitative mass results show a 412.2 ng deposition of Ag on the WE, and a 345.6 ng deposition on the CE. Separation and removal of liquid loading indi-cates an overall ~10 ng error in the magnitude of the mass change, which can reach 25 ng at certain parts of the CV. The versatility of the dual EC-QCM setup is further demonstrated by performing measurements of the third harmonic component, indicating a good agreement with the first harmonic data. Frequency tracking of higher harmonics of the motional resonance frequency is favorable for cases involving particularly small mass deposition. We also note that the presented approaches are readily applicable to electrochemical systems where parallel monitoring of mass deposition and/or liquid loading of two electrodes is required. The introduced technique is especially interesting for Li-ion battery applications where characterization of aging and degradation processes are important to investigate in real or close to real conditions by monitoring multiple QCMs.